\documentclass{article}%
\usepackage{amsmath}
\usepackage{amsfonts}
\usepackage{amssymb}
\usepackage{graphicx}%
\usepackage{tikz}
\usepackage{color}
\usepackage{cite}
\usepackage{multirow}%
\usepackage{graphicx}%
\usetikzlibrary{matrix}
\makeindex
\begin{document}
\title{\bf M2 to D2 and  vice versa  by 3-Lie and  Lie bialgebra  }
\author { M.Aali-Javanangrouh$^{a}$
\hspace{-2mm}{ \footnote{ e-mail:
aali@azaruniv.edu}}\hspace{1mm},\hspace{1mm} A. Rezaei-Aghdam $^{a}$ \hspace{-2mm}{ \footnote{Corresponding author. e-mail:rezaei-a@azaruniv.edu }}\hspace{2mm}\\
{\small{\em
$^{a}$Department of Physics, Faculty of science, Azarbaijan Shahid Madani University }}\\
{\small{\em  53714-161, Tabriz, Iran  }}}
\maketitle
\begin{abstract}

Using the concept of 3-Lie bialgebra, which has recently been defined in arXiv:1604.04475, we construct Bagger-Lambert-Gustavson (BLG) model for M2-brane on Manin triple of a special 3-Lie bialgebra. Then by  using of   correspondence  and  relation between those 3-Lie bialgebra with Lie bialgebra, we reduce this model to an $N=(4,4)$ WZW model (D2-brane), such that, its algebraic structure is a Lie bialgebra with one 2-cocycle. In this manner by using  correspondence of 3-Lie bialgebra and Lie bialgebra (for this special 3-Lie algebra) one can construct M2-brane from a D2-brane and vice versa.
\end{abstract}
\section{\label{sec:intro}Introduction}
M-theory is a magical theory with quite little knowledge about it, such that, to improve its current definition we must focus on 11 dimensional (11d) supergravity \cite{CS}. These supergravity theories are low-energy limit of M-theory, whereas, 10d supergravity is the low-energy limit  of superstring. Therefore, there should be a connection between M-theory and string theory like existing connection between  11d and 10d  supergravity. If we could make this connection clear, then a lot of unknown issues about M-theory would be resolved. Our knowledge about M-theory has been achieved by compare and grope with D-branes \cite{PK}. Many attempts have been made to obtain effective action for multiple M2-brane, the most important of which are cited in \cite{Bag1,Bag2,Bag3,Gus,ABJM,Bag4}. Basu and Harvey \cite{Bas} applied 3-bracket in BPS equation in order to explain N coincided M2-branes ending on M5-brane. This equation has been given by comprising Nahm equation for string theory \cite{Nam}. Bagger and Lambert \cite{Bag1,Bag2,Bag3} and Gustavsson \cite{Gus} independently write transformation of fields for M2-brane according to D2-brane. They obtain equation of motion of fields by using closure of  supersymmetric transformation algebra, and writing  a Lagrangian in a way that its equations of motion are the same. Bagger-Lambert-Gustavsson (BLG) model  \cite{Bag1,Bag2,Bag3,Gus} has a Lagrangian with maximal supersymmetry ($N=8$)  for description of two M2-branes \cite{Raam}, which use 3-Lie algebra. As the earlier example,  the algebra $A_4$ was the only known non-trivial  3-Lie algebra \cite{Bag3}.  Mukhi and Papageorgakis \cite{Muk2} were able to convert the topological term (Chern-Simon)  to the dynamical one, i.e, Yang-Mills term, by assigning a vacuum expectation value to a scalar fields of BLG lagrangian and  using Higgs mechanism. On the other hand, if one dualizes  M-theory on a circle, one can obtain type IIA string theory (D2-brane) \cite{Shamik}.  This can be considered as a trick for going from M2 to D2, as it was performed for $A_4$. Consequently,  a 3-Lie algebra was constructed from arbitrary Lie algebra \cite{Mat} and  BLG model on this 3-Lie algebras were studied in later works \cite{Gom,Pw,Li,MF,Paul}. In all these works one can obtain M2-model to D2 but these are not standard methods of construction of M2 from D2. Here we will try to perform another method using the concept of 3-Lie bialgebra \cite{GR}.
\par
Lie bialgebras \cite{Drinfeld} are algebraic structures of  Poisson-Lie groups \cite{vaizman} which play an important role in the theory of classical integrable systems (see \cite{Sch} for a review).  They also play an important role in $N=(2,2)$ and $N=(4,4)$ supersymmetric WZW models \cite{ Par2,Lin}. In Ref. \cite{AR} we have studied the algebraic structure of  $N=(2,2)$ and $N=(4,4)$ supersymmetric WZW models in more detail. The concept of 3-Lie algebra was described in Fillipov's work for the first time \cite{Fil} following the pioneering work of Nambu in different formulation of classical mechanics \cite{Nambu}. In Ref. \cite{GR} we define the concept of 3-Lie bialgebra by using  cohomology of 3-Lie algebras. We believe that introduction of 3-Lie bialgebras can play an important role in  M-theory. In this paper we will follow some steps in this direction. We will express the BLG Lagrangian on the Manin triple of a especial 3-Lie bialgebra  and use similar procedure applied by Mukhi and Papageorgakis \cite{Muk2}, in obtaining the  Yang-Mills in addition to other terms which are that of WZW. The extra term is square of  B-field for WZW models \cite{PS}. If the space-time coordinates were algebraic indices (like space-time coordinates ,i.e., scalar and fermion fields in BLG model ) then B-field of $N=(4,4)$ WZW model could be obtained from this form. In this way if one knows about $N=(4,4)$ WZW models in detail, then  one can obtain information about  BLG and vice versa, i.e., one can construct M2 from D2-model and vice versa.
\par
 The outline of the paper is as follows. We review BLG action and the correspondence  between M-theory and string theory in section two. In section three we review the definition of  3-Lie bialgebra \cite{GR} and give an example ,such that, is  a one-to-one correspondence between 3-Lie bialgebra and Lie bialgebra. Consequently,  in section four we express BLG model (M2-model) on the Manin triple of that 3-Lie bialgebra  and show that it turns into  Yang-Mills and  $N=(4,4)$ WZW model (D2-model). In this manner we show that using the correspondence of 3-Lie bialgebra and Lie bialgebra one can construct M2-model from D2  and vice versa.  For further description of the obtained model; in the section five we show that the WZW model can be obtained from a DBI action with extra Lie algebra valued fields.

\section{\label{sec:BLG}BLG model}
Here for self consistency of the paper and presentation of the notation, we give a short  review of  BLG model. The multiple M2-brane model of Bagger-Lambert \cite{Bag1,Bag2,Bag3} and Gustavsson \cite{Gus} (BLG) is based on 3-Lie algebra. In this algebra the Lie bracket is generalized to 3-Lie bracket \cite{Fil}. n-Lie algebra was introduced by Filippov in 1985 \cite{Fil} as an extension of the Nambu bracket \cite{Nambu} to Lie algebras. 3-Lie algebras are a special kind of n-Lie algebras, and have many applications in  mathematical and theoretical physics \cite{Bag1,Bag2,Bag3,Gus}.
The 3-Lie algebra $\cal A$ \cite{Fil},\cite{takh} with the basis $\{T^a\}$ is a vector space $\cal A$ which endowed  with the following three antisymmetric  bracket:
\begin{eqnarray}
    [T^a,T^b,T^c]=f^{abc}\hspace{0cm}_d T^d,\hspace{1cm}a,b,c,d=1,...,dim{\cal A},
  \label{3-bracket}
\end{eqnarray}
such as, to  satisfy the following fundamental identity \cite{Fil}:
\begin{eqnarray}
[T^g,T^d,[T^a,T^b,T^c]]=
[[T^g,T^d,T^a],T^b,T^c]+[T^a,[T^g,T^d,T^b],T^c]
+[T^a,T^b,[T^g,T^d,T^c]],
\label{fundamental identity}
\end{eqnarray}
where it can be redefined  by   structure constant of  ${\cal A}$ ($f^{abc}_d$ )  in the following form:
\begin{eqnarray}
f^{abc}\hspace{0cm}_ef^{gde}\hspace{0cm}_f- f^{gda}\hspace{0cm}_ef^{ebc}\hspace{0cm}_f-f^{gdb}\hspace{0cm}_ef^{aec}\hspace{0cm}_f-f^{gdc}\hspace{0cm}_ef^{abe}\hspace{0cm}_f=0.
\end{eqnarray}
\\
For the BLG model the following supersymmetric (SUSY) transformations are proposed\cite{Bag1,Bag2,Bag3}, as
\begin{eqnarray}
\label{SUSY}
 \delta X^I_a&=& i\bar{\epsilon} \Gamma^I \Psi_a,
 \nonumber
\\
 \delta\Psi_a&=& D_\mu X^I_a \Gamma^\mu \Gamma_I \epsilon -\frac{1}{2}X^I_bX^J_cX^K_d f^{bcd}\hspace{0cm}_a \Gamma_{IJK}\epsilon ,
 \nonumber
\\
 \delta(\hat{A}_\mu)_b^a&=&i\bar{\epsilon} \Gamma_\mu\Gamma_I X^I_c \Psi_d f^{cda}\hspace{0cm}_b,
\label{Supersymmetric transformation}
\end{eqnarray}
where $(\hat{A}_{\nu})^b_a=f^{cdb}\hspace{0cm}_a A_{\nu cd}$, and indices $I,J,...=1,2,...,8$ apply for transverse coordinates with $SO(8)$ symmetry (R-symmetry); and indices $\mu, \nu, ...=0,1,2$  indicate the world volume coordinate with symmetry $SO(1,2)$\footnote{Presence of M2-brane breaks Lorentz invariance $SO(1,10)$ to $ SO(1,2)\times SO(8) $.}.  Also, $\Gamma_I$s are Dirac matrices and $X^{I}$  are the transverse coordinates  of 3-Lie algebra valued coordinates and $\Psi$ is a 16 component Majorana spinor of 3-Lie algebra valued, conforming  with chirality condition using the following relation:
\begin{equation}
\Gamma^{012}\Psi=-\Psi,
\end{equation}
such that, for supersymmetric parameter $\epsilon$ we have:
\begin{equation}
\Gamma^{012}\epsilon=\epsilon
,
\end{equation}
and the covariant derivative $ D_{\mu} $ has the form :
\begin{equation}
D_{\mu}X^{(I)}_a=\partial_{\mu}X^{(I)}_a+f^{bcd}\hspace{0cm}_aA_{\mu cd}X^{(I)}_b.
\end{equation}
Furthermore, the $ \Gamma_{IJ} $ and $ \Gamma_{IJK} $ have the following forms:
\begin{eqnarray}
\Gamma^{IJ}=\frac{1}{2}(\Gamma^I\Gamma^J- \Gamma^J\Gamma^I),
\\
\{\Gamma^I,\Gamma_{JKL}\}=6\delta^I_{[J}\Gamma_{KL]}.
\end{eqnarray}
\\
 Using the assumption that the algebra of SUSY transformations  (\ref{Supersymmetric transformation})  must be closed,
 the following relations can be realized \cite{Bag1}:
 \begin{eqnarray}
 \label{mption si}
&& \Gamma^\mu D_\mu\Psi_a + \frac{1}{2}\Gamma_{IJ} X^I_cX^J_d\Psi_b f^{cdb}\hspace{0cm}_a=0,
 \\
  \label{mption X}
 &&D^2X^I_a -\frac{i}{2}\tilde{\Psi}_c\Gamma^I_JX^J_d\Psi_b f^{cdb}\hspace{0cm}_a+ \frac{1}{2}f^{bcd}\hspace{0cm}_af^{efg}\hspace{0cm}_d X^J_bX^K_cX^I_eX^J_fX^K_g=0,
 \\
  \label{mption F}
 &&(\hat{F}_{\mu\nu})^b_a + \epsilon_{\mu\nu\lambda}(X^J_cD^\lambda X^J_d +\frac{i}{2}\bar{\Psi}_c \Gamma^\lambda\Psi_d)f^{cdb}\hspace{0cm}_a=0,
  \end{eqnarray}
where,
\begin{eqnarray}
(\hat{F}_{\mu\nu})^b_a=\partial_{\mu}(\hat{A_{\nu}})^b_a-\partial_{\nu}(\hat{A_{\mu}})^b_a+(\hat{A}_{\mu})^b_c(\hat{A}_{\nu})^c_a-(\hat{A}_{\nu})^b_c(\hat{A}_{\mu})^c_a,
\end{eqnarray}
and $D^2=D_{\mu}D^{\mu}$. Similarly Bagger and Lambert have proposed the following Lagrangian \cite{Bag2} such that  the relations (\ref{mption si})-(\ref{mption F}) are its equations of motion:
\begin{eqnarray}
\nonumber
L=&-&\frac{1}{2}D_{\mu}X^{a(I)}D^{\mu}X_{a}^{(I)}+\frac{i}{2}\bar{\psi}^{a}\Gamma^{\mu}D_{\mu}\psi_{a}+\frac{i}{4}f^{abcd}\bar{\psi}_{b}\Gamma^{IJ}X_{c(I)}X_{d(J)}\psi_{a}
\\
\nonumber
&-&\frac{1}{12}f^{abcd}{f^{efg}}_{d}X_{a}^{(I)}X_{b}^{(J)}X_{c}^{(K)}X_{e}^{(I)}X_{f}^{(J)}X_{g}^{(K)}
\\
&+&\frac{1}{2}\epsilon^{\mu\nu\lambda}[f^{abcd}{A_{\mu ab}}\partial_{\nu}{A_{\lambda cd}}+\frac{2}{3}{f^{cda}}_{g}f^{efgb}{A_{\mu ab}}{A_{\nu cd}}{A_{\lambda ef}}].
\label{BLG  action}
\end{eqnarray}
The above Lagrangian is invariant under SUSY transformation  (\ref{Supersymmetric transformation}). In order for the degrees of Fermion and Boson not to vary in Lagrangian, one must use topological term which is the Chern-Simon term (the fifth  term in the bracket at the above Lagrangian) \cite{Schwars}:
\begin{eqnarray}
L_{CS}=Tr(\epsilon^{\mu\nu\lambda}(A_{\mu}\partial_{\nu}A_{\lambda}+\frac{2i}{3}(A_{\mu}A_{\nu}A_{\lambda}))),
\end{eqnarray}
with
\begin{eqnarray}
F_{\mu\nu}=\partial_{\mu}A_{\nu}-\partial_{\nu}A_{\mu}+i[A_{\mu},A_{\nu}],
\end{eqnarray}
which equipped with 3-Lie algebra. Now, we will try to extract the BLG model from a superstring model  (D2) and vice versa. In our perspective for this propose we need to apply the concept of 3-Lie bialgebra. The definition of 3-Lie bialgebra is given in \cite{GR}, however, for self containment of the paper we give a short review of this concept in the following section.

\section{\label{sec:3-Lie bialgebra1}3-Lie bialgebra}
In this section we review  the definitions of 3-Lie bialgebra.
\\

{\bf Definition}: A  Lie algebra ${ \cal G}$ with co commutator $\delta : { \cal G} \rightarrow
{ \cal G}\otimes{ \cal G} $ is a Lie bialgebra if  \cite{Sch}:
\\
{\it a} ) $\delta$ is a one-cocycle, i.e.:
\begin{eqnarray}
\delta([T^{i},T^{j}])&=&{ad^{(2)}}_{T^j}\delta(T^{i})-{ad^{(2)}}_{T^i}\delta(T^{j}),
\end{eqnarray}
where
\begin{eqnarray}
{ad^{(2)}}_{T^j}&=&ad_{T^j}\otimes 1+1\otimes ad_{T^j},
\end{eqnarray}
and $\{T^i\}$s are  bases for the Lie algebra ${ \cal G}$,
(here 1 is an identity map on $\cal G$),\\
{\it b} ) the dual map $ ^t\delta:{ \cal G}^*\otimes{ \cal G}^* \rightarrow  { \cal G}^*$ is a commutator on ${ \cal G}^*$(dual space of ${ \cal G}$) as the following definition:
\begin{eqnarray}
 (\tilde{T}_{i}\otimes \tilde{T}_{j},\delta(T^k))=({^t\delta(\tilde{T}_{i} \otimes \tilde{T}_{j}),T^k)=([\tilde{T}_{i},\tilde{T}_{j}]},T^k),
\end{eqnarray}
where $ \{{\tilde T}_i\} $ is the base for the space $ {\cal G}^* $ and
$(,)$ is the pairing between ${ \cal G}$ and ${ \cal G}^*$ . In this way there is a Lie algebra structure on the space ${ \cal G}^*$. The Lie bialgebra is shown with $({ \cal G},\delta)$ or $({ \cal G},{ \cal G}^*)$.
\\

{\bf Definition}:
$({\cal D},{ \cal G},{ \cal G}^*) $ is a Manin triple, the triple of Lie algebras $\cal D$, $\cal G$ and ${\cal G}^*$, such that,  there is a nondegenerate, symmetric and ad-invariant inner product on ${\cal D}$ with the following properties  \cite{Sch}:
\\
a) ${ \cal G}$ and ${ \cal G}^*$ are subalgebras of  $ {\cal D} $.
\\
b) ${\cal D}={ \cal G}\oplus { \cal G}^*$ as a vector space.
\\
c) ${ \cal G}$ and ${ \cal G}^*$ are isotropic, i.e.,
\begin{eqnarray}
\nonumber
(T^i,{\tilde T}_j)=\delta^i_j,\hspace{1cm}(T^i,T^j)=({\tilde T}_i,{\tilde T}_j)=0.
\end{eqnarray}
\\
The Jacobi identity for ${\cal D}={ \cal G}\oplus { \cal G}^*$ results in the following identities \cite{RH}:
\begin{eqnarray}
f^{ij}\hspace{0cm}_kf^{kl}\hspace{0cm}_m-f^{ik}\hspace{0cm}_mf^{jl}\hspace{0cm}_k+f^{jk}\hspace{0cm}_mf^{il}\hspace{0cm}_k=0,
\label{jacobian identity1}
\\
{\tilde f}_{ij}\hspace{0cm}^k{\tilde f}_{kl}\hspace{0cm}^m-{\tilde f}_{ik}\hspace{0cm}^m{\tilde f}_{jl}\hspace{0cm}^k+{\tilde f}_{jk}\hspace{0cm}^m{\tilde f}_{il}\hspace{0cm}^k=0,
\label{jacobian identity2}
\\
-f^{ij}\hspace{0cm}_k \tilde{f}_{lm}\hspace{0cm}^k+f^{ik}\hspace{0cm}_l \tilde{f}_{km}\hspace{0cm}^j-f^{jk}\hspace{0cm}_m \tilde{f}_{lk}\hspace{0cm}^i-f^{jk}\hspace{0cm}_l \tilde{f}_{km}\hspace{0cm}^i+f^{ik}\hspace{0cm}_m \tilde{f}_{lk}\hspace{0cm}^j=0,
\label{mix jacobian identity}
\end{eqnarray}
where $f^{ij}\hspace{0cm}_k$ and ${\tilde f}_{ij}\hspace{0cm}^k$ are the structure constants of the Lie algebras $\cal G$ and ${ \cal G}^*$, respectively   (i.e. $[T^i,T^j]=f^{ij}\hspace{0cm}_kT_k$ ,$[{\tilde T}_i,{\tilde T}_j]={\tilde f}_{ij}\hspace{0cm}^k{\tilde T}_k$). Note that  (\ref{jacobian identity1}) and (\ref{jacobian identity2}) are Jacobi identities for the Lie algebras $\cal G$ and $ \cal G^* $, respectively, and (\ref{mix jacobian identity}) is the mix Jacobi identity on ${\cal D}$.
\\
\par
{\bf Theorem}: {\em There exist a one to one correspondence between Lie bialgebra $({ \cal G},{ \cal G}^*)$ and Manin triple $({\cal D},{ \cal G},{ \cal G}^*)$}\cite{Sch}.
\\
\par
 We have defined the 3-Lie bialgebra  in  \cite{GR} as follows:
\\
\par
{\bf Definition}: A  3-Lie algebra ${ \cal A}$ with co commutator $\delta : { \cal A} \rightarrow
{ \cal A}\otimes{ \cal A} \otimes{ \cal A} $ is a  \emph{3-Lie bialgebra} if  \cite{GR}:

{\it a} ) $\delta$ is a one-cocycle of ${\cal A}$ with value in $ \otimes^3 {\cal A} $, i.e:
\begin{eqnarray}
\label{one cocycle}
\delta([T^{a},T^{b},T^{c}])&=&{ad^{(3)}}_{T^{b}\otimes T^{c}}\delta(T^{a})-{ad^{(3)}}_{T^{a}\otimes T^{c}}\delta(T^{b})+{ad^{(3)}}_{T^{a}\otimes T^{b}}\delta(T^{c}),
\end{eqnarray}
such that,
\begin{eqnarray}
{ad^{(3)}}_{T^{b}\otimes T^{c}}&=&ad_{T^{b}\otimes T^{c}}\otimes 1\otimes 1+1\otimes ad_{T^{b}\otimes T^{c}}\otimes1+1\otimes1\otimes ad_{T^{b}\otimes T^{c}},
\end{eqnarray}
\\
where $\{T^a\}$s are  bases of 3-Lie algebra ${\cal A}$ and we have $ ad_{T^a \otimes T^b}T^c=[T^a,T^b,T^c] $\cite{Taktajan}.
\\
{\it b} ) the dual map $ ^t\delta:\otimes^3 {\cal A}^* \rightarrow  {\cal A}^*$ is a 3-Lie bracket on ${\cal A}^*$ (dual space of ${\cal A}$ ) which is a commutator on $ \cal G^* $ satisfying  the fundamental identity :
\begin{equation}
 (\tilde{T}_{a}\otimes \tilde{T}_{b} \otimes \tilde{T}_{c},\delta(T^d))=({^t\delta(\tilde{T}_{a} \otimes \tilde{T}_{b}\otimes \tilde{T}_{c}),T^d)=([\tilde{T}_{a},\tilde{T}_{b},\tilde{T}_{c}]},T^d),
 \label{identity}
\end{equation}
in which $\{{\tilde T}_a\}$ is the base for the space ${ \cal A}^*$ and  $( , )$ is a natural pairing between ${\cal A}$ and $ {\cal A}^*$. In this way ${ \cal A}^*$ constructs a 3-Lie algebra. The 3-Lie bialgebra can be denoted either by $({\cal A} ,{\cal A}^*)$ or $({\cal A},\delta )$.
 \\
\par
{\bf Definition}: $({\cal D},{ \cal A},{ \cal A}^*) $ is Manin triple, a triple of 3-Lie algebras $\cal D$, $\cal A$ and ${\cal A}^*$ such that there is a nondegenerate, symmetric and ad-invariant inner product on ${\cal D}$ with the following properties \cite{GR}:
\footnote{
Note that in general  vector space $\cal D$  is not a 3-Lie algebra.}
\\
a) ${ \cal A}$ and ${ \cal A}^*$ are 3-Lie subalgebras of $ {\cal D} $,
\\
b) ${\cal D}={ \cal A}\oplus { \cal A}^*$ as a vector space,
\\
c) ${ \cal A}$ and ${ \cal A}^*$ are isotropic, i.e. \begin{eqnarray}
\nonumber
(T^a,{\tilde T}_b)=\delta^a_b,\hspace{1cm}(T^a,T^b)=({\tilde T}_a,{\tilde T}_b)=0.
\end{eqnarray}
\\
 By using the fundamental identity (\ref{fundamental identity}), equation  (\ref{one cocycle}) and relation $ \delta(T^{a})=\tilde{f}_{ bcd}\hspace{0cm}^a T^{b}\otimes T^{c} \otimes T^{d},
  \label{delta} $ one can obtain the following fundamental and  mix fundamental identities \cite{GR}:
\begin{eqnarray}
f^{aef}\hspace{0cm}_gf^{bcdg}&-& f^{bef}\hspace{0cm}_gf^{acdg}+f^{cef}\hspace{0cm}_gf^{abdg}-f^{def}\hspace{0cm}_gf^{abcg}=0,
\\
{\tilde f}_{aef}\hspace{0cm}^g{\tilde f}_{bcdg}&-& {\tilde f}_{bef}\hspace{0cm}^g{\tilde f}_{acdg}+ {\tilde f}_{cef}\hspace{0cm}^g{\tilde f}_{abdg}-{\tilde f}_{def}\hspace{0cm}^g{\tilde f}_{abcg}=0,
\\
\nonumber
{f^{abc}}_{g}\:{\tilde{f}_{def}\:}^{g}&=&{f^{gbc}}_{f}\:{\tilde{f}_{deg}\:}^{a}+{f^{gbc}}_{e}\:{\tilde{f}_{dfg}\:}^{a}
-{f^{gbc}}_{d}\:{\tilde{f}_{efg}\:}^{a}-{f^{gac}}_{f}\:{\tilde{f}_{deg}\:}^{b}+{f^{gac}}_{e}\:{\tilde{f}_{dfg}\:}^{b}
\\
&-&{f^{gac}}_{d}\:{\tilde{f}_{efg}\:}^{b}+{f^{gab}}_{f}\:{\tilde{f}_{deg}\:}^{c}-{f^{gab}}_{e}\:{\tilde{f}_{dfg}\:}^{c}+{f^{gab}}_{d}\:{\tilde{f}_{efg}\:}^{c},
\label{4-4}
\end{eqnarray}
where $f^{abc}\hspace{0cm}_d$ and $\tilde{f} _{abc}\hspace{0cm}^d$ are structure constants of 3-Lie algebras ${\cal A}$ and ${\cal A}^*$, respectively.

\subsection{\label{An example}An example}
Now, we will consider  an especial  example 3-Lie bialgebra  related to 3-Lie algebra ${\cal A}_{ \cal G}$ and Lie algebra ${\cal G}$\footnote{Note that this example of 3-Lie bialgebra was considered in \cite{Aali} as a first time. Here for a self containing of the paper we denote it as an example.}.
The 3-Lie algebras ${\cal A}_{ \cal G}$ (mentioned in \cite{Mat} for a first time) have commutation relations as  follows:
\begin{eqnarray}
\label{example 3-bracket}
[T^{-},T^{a},T^{b}]=0,\hspace{1cm}
[T^{+},T^{i},T^{j}]={f}^{ij}\hspace{0cm}_{k}T^{k},\hspace{1cm}
[T^{i},T^{j},T^{k}]=f^{ijk}T^{-},
\end{eqnarray}
where $\{T^i\}$s  are basis  of the Lie algebra ${ \cal G}$ ($[T^{i},T^{j}]={f}^{ij}\hspace{0cm}_{k}
T^{k}$ with $i,j,k=1,2,...,dim { \cal G}$) and $f^{ij}\hspace{0cm}_k$ is its structure constant\footnote{Note that the indices of $f^{ij}\hspace{0cm}_k$ are lowered and raised by the ad-invariant metric  $g^{ij}$ of the Lie algebra $\cal G.$}. Furthermore, $T^{-}$ and $T^+$ are new generators and we have $a=+,-,i$.
Now we propose that  there exists a  3-Lie algebra structure on ${\cal A}^*_{\cal G^*}$ with similar  commutation relations:
\begin{eqnarray}
\label{2example 3-bracket}
[\tilde{T}_{-},\tilde{T}_{a},\tilde{T}_{b}]=0,\hspace{1cm}
[\tilde{T}_{+},\tilde{T}_{i},\tilde{T}_{j}]={\tilde{f}}_{ij}\hspace{0cm}^{k}\tilde{T}_{k},\hspace{1cm}
[\tilde{T}_{i},\tilde{T}_{j},\tilde{T}_{k}]=\tilde{f}_{ijk}T_{-},
\end{eqnarray}

such that  ${\cal G}^*$($[\tilde{T}_{i},\tilde{T}_{j}]={\tilde{f}}_{ij}\hspace{0cm}^{k}\tilde{T}^{k}$ with $i,j,k=1,2,...,dim { \cal G^*}$) is a Lie algebra.
\par
 {\bf Proposition: } \cite{Aali} A 3-Lie algebra ${\cal A}_{\cal G}$ construct a 3-Lie bialgebra $({\cal A}_{\cal G},{\cal A}_{\cal G}^*)$ if and only if $({\cal G},{\cal G}^*)$is a Lie bialgebra. The proof can be found in \cite{Aali}.

\section{\label{BLG model on Manin triple of 3-Lie algebra }BLG model on Manin of 3-Lie algebras ( M2 $\leftrightarrow$ D2 ) }
In the previous section we have considered a especial case of the Manin triple  $({\cal D}, {\cal A}_{\cal G},{\cal A}_{{\cal G}^*})$ and have noted that  there is a correspondence between  3-Lie bialgebra  $({\cal A}_{\cal G},{\cal A}_{{\cal G}^*})$ and Lie bialgebra $({\cal G},{\cal G}^*)$\footnote{In general the vector space $\cal D$ in  triple $({\cal D}, {\cal A}_{\cal G},{\cal A}_{{\cal G}^*})$  is not 3-Lie algebra and also  there is not correspondence between Manin triple  $({\cal D}, {\cal A}_{\cal G},{\cal A}_{{\cal G}^*})$ and Lie bialgebra $({\cal G},{\cal G}^*)$, but for the above  special example ${\cal D}$  is a 3-Lie algebra and for  this case there is correspondence.}. Now we want to apply this 3-Lie algebra $\cal D$ in the  BLG model.
We obtained in the  previous section that:
\begin{eqnarray}
F^{-AB}\hspace{0cm}_{C}=0,\hspace{1cm}F^{{\tilde-}AB}\hspace{0cm}_{C}=0,
\end{eqnarray}
then
\begin{eqnarray}
F^{ABC}\hspace{0cm}_{+}=0,\hspace{1cm}F^{ABC}\hspace{0cm}_{{\tilde+}}=0,
\end{eqnarray}
 Note that for the Manin triple ${\cal D}$ we  apply the symbol $ F^{ABC}\hspace{0cm}_D $ for the structure constant of the Manin triple as a $(4+2 dim{{\cal G}})$\footnote{We assume ${\cal {G}}$ and ${\cal{G}^*}$ have same dimension i.e. $2+ dim{\cal G}$ that 2 is used for $+$ and $-$.} dimensional 3-Lie algebra, i.e. we have ${T^A}$ as a basis for the Manin triple with $A=+$, $T^+=T^+$; $A=-$, $T^-=T^-$; $A=i$, $T^i=T^i$; $A= i+ (2+dim{\cal G})$, $T^{ i+(2+ dim{\cal G})}={T}^{\tilde i}$; $A= (-)+(2+dim{\cal G})$, $T^{ (-)+(2+dim{\cal G})}=T^{\tilde -}$ and $A=(+)+(2+dim{\cal G})$, $T^{ (+)+(2+dim{\cal G})}=T^{\tilde +}$ together with the following commutation relations:
\begin{eqnarray}
\nonumber
&&[T^-,T^A,T^B]=0,\hspace{0.5cm}[T^+,T^i,T^j]=f^{ij}\hspace{0cm}_k T^k,\hspace{0.5cm}[T^+,T^i,T^{\tilde{j}}]= f^{ik}\hspace{0cm}_{j}T^{\tilde{k}},\hspace{0.5cm}[T^i,T^j,T^k]=f^{ijk}T^-,
\\
\nonumber
&&[T^{\tilde{-}},T^A,T^B]=0, \hspace{0.5cm}[T^{\tilde{+}},T^{\tilde{i}},T^{\tilde{j}}]=\tilde{f}_{ij}\hspace{0cm}^kT_{\tilde{ k}},\hspace{0.5cm}[T^{\tilde{+}},T^{\tilde{i}},T^j]={\tilde{f}}_{ik}\hspace{0cm}^j T^k,\hspace{0.5cm}[T^{\tilde{i}},T^{\tilde{j}},T^{\tilde{k}}]=\tilde{f}_{ijk}T_{\tilde{-}},
\\
\label{123}
&&\hspace{-0.5cm}[T^{\tilde{k}},T^i,T^j]=f^{ji}\hspace{0cm}_kT^{\tilde{-}},\hspace{0.5cm}[T^{\tilde{+}},T^j,T^k]=f^{ijk}T^{\tilde{i}},\hspace{0.5cm}[T^+,T^{\tilde{j}},T^{\tilde{k}}]={\tilde f}_{ijk}T^i,\hspace{0.5cm}[T^k,T^{\tilde{i}},T^{\tilde{j}}]=-{\tilde f}_{ij}\hspace{0cm}^kT^-.
\end{eqnarray}

Now we write the equations of motion for the BLG model (\ref{mption si})-(\ref{mption F}) by considering 3-Lie algebra $\cal D$ of the Manin triple related to this especial 3-Lie bialgebra as  follows:
 \begin{eqnarray}
 \label{mption manin Psi}
&& \Gamma^\mu D_\mu\Psi_A + \frac{1}{2}\Gamma_{IJ} X^I_CX^J_D\Psi_B F^{CDB}\hspace{0cm}_A=0,
 \\
  \label{mption manin X}
 &&D^2X^I_A -\frac{i}{2}\tilde{\Psi}_C\Gamma^I_JX^J_D\Psi_B F^{CDB}\hspace{0cm}_A+ \frac{1}{2}F^{BCD}\hspace{0cm}_AF^{EFG}\hspace{0cm}_D X^J_BX^K_CX^I_EX^J_FX^K_G=0,\\
 &&(\hat{F}_{\mu\nu})^B_A + \epsilon_{\mu\nu\lambda}(X^J_CD^\lambda X^J_D +\frac{i}{2}\bar{\Psi}_C \Gamma^\lambda\Psi_D)F^{CDB}\hspace{0cm}_A=0,
   \end{eqnarray}
such that, if we take $A=+,\tilde +$ in (\ref{mption manin Psi}) and (\ref{mption manin X}), then we obtain the following relations:
\begin{eqnarray}
 \label{constraint1}
\partial^2 X^I_+ = 0,
\qquad
\Gamma^{\mu}\partial_{\mu} \Psi_+ = 0,\hspace{1cm}\partial^2 X^I_{{\tilde+}} = 0,
\qquad
\Gamma^{\mu}\partial_{\mu} \Psi_{{\tilde+}} = 0,
\end{eqnarray}
this means that ($X^I_+, X^I_{\tilde+}$)and ($ \Psi_+, \Psi_{{\tilde+}}$)  can be set to a constant  such as  Yang-Mills coupling \cite{Shamik} as applied to the whole theory and zero, respectively. This constant  must conserve the SUSY transformations (\ref{SUSY}) for Manin triple. Then, from the following relations:
\begin{eqnarray}
\label{1}
\delta X^I_+ &=& i\bar{\epsilon} \Gamma^I \Psi_+,\\
\label{2}
\delta \Psi_+ &=& \partial_{\mu} X^I_+ \Gamma^{\mu}\Gamma^I \epsilon,
\end{eqnarray}
one can show that SUSY transformations do not change if we assign  a vacuum expectation value (VEV) to one of the fields.
Then, the Lagrangian terms become as follows:
by considering this relations:
 \begin{eqnarray}
D_\mu X^{(I)}_-&=&\partial_\mu X^{(I)}_-+f^{ijk}A_{\mu jk}X^{(I)}_i+2\tilde{f}_{ji}\hspace{0cm}^{k}A_{\mu \tilde{j}k}X^{(I)}_{\tilde{i}}+\tilde{f}_{kj}\hspace{0cm}^iA_{\mu \tilde{j}\tilde{k}}X^{(I)}_{i} ,
\\
D_\mu X^{(I)}_{\tilde -}&=&\partial_\mu X^{(I)}_{\tilde -}+\tilde{f}_{ijk}A_{\mu\tilde{j}\tilde{k}}X^{(I)}_{\tilde i}+f^{jk}\hspace{0cm}_iA_{\mu jk}X^{(I)}_{\tilde i}+2f^{ij}\hspace{0cm}_{k}A_{\mu j\tilde{k}}X^{(I)}_{i},
\\
\nonumber
D_\mu X^{(I)}_i&=&\partial_\mu X^{(I)}_i+f^{jk}\hspace{0cm}_iA_{\mu jk}X^{(I)}_+ +{\tilde f}_{ijk}A_{\mu {\tilde j}{\tilde k}}X^{(I)}_++2f^{jk}\hspace{0cm}_iA_{\mu k}X^{(I)}_j-2{\tilde f}_{ijk}A_{\mu \tilde k}X^{(I)}_{\tilde j}\\
&&+2{\tilde f}_{ji}\hspace{0cm}^kA_{\mu {\tilde j}k}X^{(I)}_{\tilde +}+2f_{ij}\hspace{0cm}^kA_{\mu k}X^{(I)}_{\tilde j}+2{\tilde f}_{ki}\hspace{0cm}^jA_{\mu \tilde k}X^{(I)}_j
\\
\nonumber
D_\mu X^{(I)}_{\tilde i}&=&\partial_\mu X^{(I)}_{\tilde i}+f^{ijk}A_{\mu jk}X^{(I)}_++{\tilde f}_{jk}\hspace{0cm}^i A_{\mu {\tilde j}{\tilde k}}X^{(I)}_{\tilde +}-2f^{ijk}A_{\mu k}X^{(I)}_j-{\tilde f}_{jk}\hspace{0cm}^i A_{\mu {\tilde k}}X^{(I)}_{\tilde j}
\\
&&+2f^{ki}\hspace{0cm}_j A_{\mu {\tilde j}k}X^{(I)}_+ -f^{ki}\hspace{0cm}_jA_{\mu k}X^{(I)}_{\tilde j}-2f^{ij}\hspace{0cm}_k A_{\mu \tilde k}X^{(I)}_j,
\end{eqnarray}
we will have
\begin{eqnarray}
\nonumber
D_\mu X^{(I)}_AD^\mu X^{A(I)}&=&\partial_\mu X^{(I)}_-\partial^\mu X^{-(I)}+\partial_\mu X^{(I)}_{\tilde -}\partial^\mu X^{{\tilde -}(I)}+\partial_\mu X^{(I)}_+\partial^\mu X^{+(I)}+\partial_\mu X^{(I)}_{\tilde +}\partial^\mu X^{{\tilde +}(I)}
\\&+&D_\mu X^{(I)}_iD^\mu X^{i(I)}+D_\mu X^{(I)}_{\tilde i}D^\mu X^{{\tilde i}(I)}
\end{eqnarray}
where
\begin{eqnarray}
\nonumber
D_\mu X^{(I)}_iD^\mu X^{i(I)}\hspace{-0.3cm}&=&\hspace{-0.3cm}\partial_\mu X^{(I)}_i\partial^\mu X^{i(I)}+[f^{jk}\hspace{0cm}_iA_{\mu jk}X^{(I)}_++2{\tilde f}_{ji}\hspace{0cm}^k A_{\mu {\tilde j}k}X^{(I)}_{\tilde +}+{\tilde f}_{ijk}A_{\mu {\tilde j}{\tilde k}}X^{(I)}_+-2f^{jk}\hspace{0cm}_iA_{\mu k}X^{(I)}_j
\\
\nonumber
&-&\hspace{-0.3cm}2{\tilde f}_{ijk}A_{\mu {\tilde k}}X^{(I)}_{\tilde j}+2f_{ij}\hspace{0cm}^k A_{\mu k}X^{(I)}_{\tilde j}+2{\tilde f}_{ki}\hspace{0cm}^jA_{\mu {\tilde k}}X^{(I)}_j][f^{ijk}A_{\mu jk}X^{-(I)}-{\tilde f}_{jk}\hspace{0cm}^iA_{\mu j{\tilde k}}X^{-(I)}
\\
&-&\hspace{-0.3cm}2f^{ik}\hspace{0cm}_jA_{\mu k}X^{j(I)}+{\tilde f}_{jk}\hspace{0cm}^iA{\mu {\tilde k}}X^{j(I)}+f^{ij}\hspace{0cm}_kA_{\mu \tilde k}X^{{\tilde j}(I)}-f^{ikj}A_{\mu k}X^{{\tilde j}(I)}-2f^{ij}\hspace{0cm}_kA_{\mu j{\tilde k}}X^{{\tilde -}(I)}],
\end{eqnarray}
and
\begin{eqnarray}
\nonumber
D_\mu X^{(I)}_{\tilde i}D^\mu X^{{\tilde i}(I)}\hspace{-0.3cm}&=&\hspace{-0.3cm}\partial_\mu X^{(I)}_{\tilde i}\partial^\mu X^{{\tilde i}(I)}+[f^{ijk}A_{\mu jk}X^{(I)}_++{\tilde f}_{jk}\hspace{0cm}^iA_{\mu {\tilde j}{\tilde k}}X^{(I)}_{\tilde +}+2f^{ki}\hspace{0cm}jA_{\mu {\tilde j}k}X^{(I)}_+-2f^{ijk}A_{\mu k}X^{(I)}_j
\\
\nonumber&-&\hspace{-0.3cm}{\tilde f}_{jk}\hspace{0cm}^iA_{\mu \tilde k}X^{(I)}_{\tilde j}-f^{ki}\hspace{0cm}_jA_{\mu k}X^{(I)}_{\tilde j}-2f^{ij}\hspace{0cm}_kA_{\mu {\tilde k}}X^{(I)}_j][2{\tilde f}_{ik}\hspace{0cm}^jA_{\mu j\tilde k}X^{-(I)}-2f^{jk}\hspace{0cm}_iA_{\mu jk}X^{{\tilde -}(I)}
\\
&-&\hspace{-0.3cm}{\tilde f}_{ijk}A_{\mu {\tilde j}{\tilde k}}X^{{\tilde -}(I)}-2f_{ikj}A_{\mu {\tilde k}}X^{j(I)}-2{\tilde f}_{ji}\hspace{0cm}^kA_{\mu k}X^{j(I)}+2f^{jk}\hspace{0cm}_iA_{\mu k}X^{{\tilde j}(I)}+2{\tilde f}_{ik}\hspace{0cm}^jA_{\mu {\tilde k}}X^{{\tilde j}(I)}].
\end{eqnarray}
Furthermore, the first term of CS term in BLG action (\ref{BLG  action}) turns into the following forms:
 \begin{eqnarray}
\frac{1}{2}\epsilon^{\mu\nu\lambda}
F^{ABCD}A_{\mu AB}\partial_\nu A_{\lambda CD}=2\epsilon^{\mu\nu\lambda}
F^{BCD}A_{\mu BC}\partial_\nu A_{\lambda D}+2\epsilon^{\mu\nu\lambda}
F^{{\cal B}{\cal C}{\cal D}}A_{\mu {\cal B}{\cal C}}\partial_\nu A_{\lambda {\cal D}}
\label{CS11}
\end{eqnarray}
where 
 \begin{eqnarray}
\nonumber
&&\epsilon^{\mu\nu\lambda}F^{BCD}A_{\mu B}\partial_\nu A_{\lambda CD}=\frac{1}{3}\epsilon^{\mu\nu\lambda}f^{jk}\hspace{0cm}_{i}A_{\mu}^i\partial_{\nu}A_{\lambda jk}+\frac{2}{3}\epsilon^{\mu\nu\lambda}f^{ij}\hspace{0cm}_{k}A_{\mu i}\partial_{\nu}A_{\lambda j}^k+\frac{2}{3}\epsilon^{\mu\nu\lambda}f^{ki}\hspace{0cm}_{j}A_{\mu i}\partial_{\nu}A_{\lambda \tilde{j}}^{\tilde k}
\\
\nonumber
&+&\frac{2}{3}\epsilon^{\mu\nu\lambda}f^{ij}\hspace{0cm}_{k}A_{\mu}^{\tilde i}\partial_{\nu}A_{\lambda j\tilde{k}}+\frac{2}{3}\epsilon^{\mu\nu\lambda}f^{jk}\hspace{0cm}_{i}A_{\mu}^{\tilde i}\partial_{\nu}A_{\lambda j}^{\tilde k}+\frac{1}{3}\epsilon^{\mu\nu\lambda}{\tilde f}_{jk}\hspace{0cm}^{i}A_{\mu}^{\tilde i}\partial_{\nu}A_{\lambda {\tilde j}{\tilde k}}+\frac{2}{3}\epsilon^{\mu\nu\lambda}{\tilde f}_{ik}\hspace{0cm}^{j}A_{\mu}^{i}\partial_{\nu}A_{\lambda {j}{\tilde k}}
\\
&+&\frac{2}{3}\epsilon^{\mu\nu\lambda}{\tilde f}_{jk}\hspace{0cm}^{i}A_{\mu}^{i}\partial_{\nu}A_{\lambda{\tilde k}}^j+\frac{2}{3}\epsilon^{\mu\nu\lambda}{\tilde f}_{k{\tilde i}}\hspace{0cm}^{j}A_{\mu\tilde i}\partial_{\nu}A_{\lambda {j}}^k+\frac{2}{3}\epsilon^{\mu\nu\lambda}{\tilde f}_{ij}\hspace{0cm}^{k}A_{\mu\tilde i}\partial_{\nu}A_{\lambda {\tilde j}}^{\tilde k}
\label{WZW1}
\end{eqnarray}
and 
 \begin{eqnarray}
\nonumber
\epsilon^{\mu\nu\lambda}
F^{{\cal B}{\cal C}{\cal D}}A_{\mu {\cal B}}\partial_\nu A_{\lambda {\cal C}{\cal D}}&=&\frac{2}{3}\epsilon^{\mu\nu\lambda}f^{ijk}\hspace{0cm}A_{\mu i}\partial_{\nu}A_{\lambda j}^{\tilde k}+\frac{1}{3}\epsilon^{\mu\nu\lambda}f^{jki}\hspace{0cm}_{i}A_{\mu}^{\tilde i}\partial_{\nu}A_{\lambda jk}
\\
&+&\frac{1}{3}\epsilon^{\mu\nu\lambda}{\tilde f}_{ijk}\hspace{0cm}^{i}A_{\mu}^{i}\partial_{\nu}A_{\lambda{\tilde j}{\tilde k}}+\frac{2}{3}\epsilon^{\mu\nu\lambda}{\tilde f}_{jik}\hspace{0cm}^{i}A_{\mu{\tilde i}}\partial_{\nu}A_{\lambda{\tilde k}}^j
 \end{eqnarray}
and the second term of CS term in BLG action 
 \begin{eqnarray}
\frac{1}{3}\epsilon^{\mu\nu\lambda}\hspace{-.08cm}F^{AEF}\hspace{-.08cm}_{G}\,F^{BCDG}\,A_{\mu AB}A_{\nu CD}A_{\lambda EF}\hspace{-.1cm}
=\hspace{-.1cm}-2\epsilon^{\mu\nu\lambda}\,F^{ABC} F^{EF}\hspace{-.08cm}_A A_{\mu EF}A_{\nu B}A_{\lambda C}-2\epsilon^{\mu\nu\lambda}\,F^{{\cal A}{\cal B}{\cal C}} F^{{\cal E}{\cal F}}\hspace{-.08cm}_{\cal A} A_{\mu {\cal E}{\cal F}}A_{\nu {\cal B}}A_{\lambda {\cal C}}
\label{CS22}
 \end{eqnarray}
where
 \begin{eqnarray}
 \nonumber
& &\hspace{-1cm}\epsilon^{\mu\nu\lambda}\hspace{-.09cm}F^{ABC}\hspace{-.08cm} F^{EF}\hspace{-.08cm}_A A_{\mu EF}A_{\nu B}A_{\lambda C}=\frac{1}{2}\epsilon^{\mu\nu\lambda}f^{jk}\hspace{0cm}_if^{il}\hspace{0cm}_m A_{\mu jk} A_{\nu l}A_{\lambda}^m+\frac{1}{2}\epsilon^{\mu\nu\lambda}f^{ij}\hspace{0cm}_kf^{lm}\hspace{0cm}_iA_{\mu j}^kA_{\nu l}A_{\lambda m}
\\
 \nonumber
&&\hspace{-1cm}+\epsilon^{\mu\nu\lambda}f^{ij}\hspace{0cm}_k\tilde{f}_{mi}\hspace{0cm}^lA_{\mu j{\tilde k}}A_{\nu l}A_{\lambda}^m+\frac{1}{2}\epsilon^{\mu\nu\lambda}f^{jk}\hspace{0cm}_i\tilde{f}_{lm}\hspace{0cm}^iA_{\mu jk}A_{\nu {\tilde l}}A_{\lambda}^m+\epsilon^{\mu\nu\lambda}f^{ij}\hspace{0cm}_k\tilde{f}_{il}\hspace{0cm}^mA_{\mu j}^kA_{\nu {\tilde l}}A_{\lambda m}+\epsilon^{\mu\nu\lambda}f^{ij}\hspace{0cm}_k f^{lm}\hspace{0cm}_iA_{\mu j{\tilde k}}A_{\nu}^{\tilde l}A_{\lambda m}
\\
 \nonumber
 &&\hspace{-1cm}+\frac{1}{2}\epsilon^{\mu\nu\lambda}f^{jk}\hspace{0cm}_i f^{mi}\hspace{0cm}_lA_{\mu j}^{\tilde k}A_{\nu {\tilde l}}A_{\lambda m}+\epsilon^{\mu\nu\lambda}f^{mi}\hspace{0cm}_l\tilde{f}_{ki}\hspace{0cm}^jA_{\mu j{\tilde k}}A_{\nu}^lA_{\lambda m}+\epsilon^{\mu\nu\lambda}f^{lm}\hspace{0cm}_i\tilde{f}_{kj}\hspace{0cm}^iA_{\mu{\tilde k}}^jA_{\nu l}A_{\lambda m}+\frac{1}{2}\epsilon^{\mu\nu\lambda}{\tilde f}_{jk}\hspace{0cm}^i\tilde{f}_{mi}\hspace{0cm}^lA_{\mu {\tilde j}{\tilde k}}A_{\nu l}A_{\lambda}^m
 \\
 \nonumber
&&\hspace{-1cm}+\epsilon^{\mu\nu\lambda}{\tilde f}_{ki}\hspace{0cm}^j\tilde{f}_{lm}\hspace{0cm}^iA_{\mu {\tilde j}{\tilde k}}A_{\nu{\tilde l}}A_{\lambda}^m+\epsilon^{\mu\nu\lambda}{\tilde f}_{kj}\hspace{0cm}^i\tilde{f}_{il}\hspace{0cm}^mA_{\mu{\tilde k}}^jA_{\nu{\tilde l}}A_{\lambda m}+\frac{1}{2}\epsilon^{\mu\nu\lambda}f^{im}\hspace{0cm}_l\tilde{f}_{ij}\hspace{0cm}^kA_{\mu{\tilde j}}^{\tilde k}A_{\nu {\tilde l}}A_{\lambda m}+\frac{1}{4}\epsilon^{\mu\nu\lambda}f^{lm}\hspace{0cm}_i\tilde{f}_{jk}\hspace{0cm}^iA_{\mu {\tilde j}{\tilde k}}A_{\nu l}A_{\lambda}^{\tilde m}
 \\
\nonumber
&&\hspace{-1cm}+\epsilon^{\mu\nu\lambda}f^{mi}\hspace{0cm}_l\tilde{f}_{ki}\hspace{0cm}^jA_{\mu j{\tilde k}}A_{\nu{\tilde l}}A_{\lambda}^{\tilde m}+\frac{1}{2}\epsilon^{\mu\nu\lambda}{\tilde f}_{jk}\hspace{0cm}^i\tilde{f}_{mi}\hspace{0cm}^lA_{\mu {\tilde j}{\tilde k}}A_{\nu}^{\tilde l}A_{\lambda {\tilde m}}+\frac{1}{2}\epsilon^{\mu\nu\lambda}{\tilde f}_{ij}\hspace{0cm}^k\tilde{f}_{lm}\hspace{0cm}^iA_{\mu {\tilde j}}^{\tilde k}A_{\nu{\tilde l}}A_{\lambda {\tilde m}}+\frac{1}{2}\epsilon^{\mu\nu\lambda}f^{jk}\hspace{0cm}_i\tilde{f}_{li}\hspace{0cm}^mA_{\mu j}^{\tilde k}A_{\nu l}A_{\lambda m}
\\
&+&\frac{1}{2}\epsilon^{\mu\nu\lambda}f^{jk}\hspace{0cm}_i f^{mi}\hspace{0cm}_lA_{\mu jk}A_{\nu{\tilde l}}A_{\lambda}^{\tilde m}+\epsilon^{\mu\nu\lambda}f^{ij}\hspace{0cm}_k\tilde{f}_{il}\hspace{0cm}^mA_{\mu j{\tilde k}}A_{\nu{\tilde l}}A_{\lambda}^{\tilde m}+\frac{1}{2}\epsilon^{\mu\nu\lambda}f^{jk}\hspace{0cm}_i\tilde{f}_{lm}\hspace{0cm}^iA_{\mu j{\tilde k}}A_{\nu{\tilde l}}A_{\lambda {\tilde m}},
 \label{WZW2}
   \end{eqnarray}
 and 
 \begin{eqnarray}
 \nonumber
&&\hspace{-1cm}\epsilon^{\mu\nu\lambda}\,F^{{\cal A}{\cal B}{\cal C}} F^{{\cal E}{\cal F}}\hspace{0cm}_{\cal A} A_{\mu {\cal E}{\cal F}}A_{\nu {\cal B}}A_{\lambda {\cal C}}=\frac{1}{3}\epsilon^{\mu\nu\lambda}{\big (}-[
2f^{jk}\hspace{0cm}_if^{il}\hspace{0cm}_m+\frac{3}{2}f^{jki}{\tilde f}_{mi}\hspace{0cm}^l+3f^{lji}{\tilde f}{mi}\hspace{0cm}^k] A_{\mu jk} A_{\nu l}A_{\lambda}^m
\\
\nonumber
&&\hspace{-1cm}+[2f^{kj}\hspace{0cm}_if^{li}\hspace{-.08cm}_m-3f^{lki}\tilde{f}_{mi}\hspace{0cm}^j-2f^{jil}\tilde{f}_{mi}\hspace{0cm}^k]A_{\mu {\tilde m}}A_{\nu k}^{\tilde j}A_{\lambda l}-[f^{jk}\hspace{-.08cm}_if^{im}\hspace{-.08cm}_l-2f^{ijm}\tilde{f}_{il}\hspace{-.08cm}^k+{\tilde f}_{il}\hspace{0cm}^mf^{jki}]A_{\mu}^{\tilde m}A_{\nu jk}A_{\lambda \tilde l}
\\
\nonumber
&&\hspace{-1cm}-2[2f^{ki}\hspace{-.08cm}_j\tilde{f}_{li}\hspace{-.08cm}^m-\frac{1}{2}f^{ikm}\tilde{f}_{ilj}+2f^{im}\hspace{-.08cm}_l\tilde{f}_{ji}\hspace{-.08cm}^k-\frac{1}{2}f^{mk}\hspace{-.08cm}_i\tilde{f}_{lj}\hspace{-.08cm}^i
-\frac{1}{2}f^{ki}\hspace{-.08cm}_l\tilde{f}_{ij}\hspace{-.08cm}^m-\frac{1}{2}f^{kim}\tilde{f}_{ilj}]A_{\mu}^{\tilde m}A_{\nu{\tilde j}k}A_{\lambda \tilde l}+3f^{il}_m\tilde{f}_{kij}A_{\mu k}A_{\nu \tilde j}A_{\lambda l\tilde m}
\\
\nonumber
&&\hspace{-1cm}+[f^{jki}\tilde{f}_{im}\hspace{-.08cm}^l+f^{lki}\tilde{f}_{mi}\hspace{-.08cm}^j+2f^{jk}_if^{li}\hspace{-.08cm}_m]A_{\mu k}^mA_{\nu }^{\tilde j}A_{\lambda l}-[\tilde{f}_{jk}\hspace{0cm}^i\tilde{f}_{il}\hspace{0cm}^m+\tilde{f}_{mk}\hspace{-.08cm}^i\tilde{f}_{ilm}]A_{\mu \tilde j}A_{\nu\tilde j}^kA_{\lambda \tilde l}+\tilde{f}_{jik}f^{im}\hspace{-.08cm}_lA_{\mu  m}^{\tilde k}A_{\nu}^{ j}A_{\lambda  l}
 \\
 \nonumber
&&\hspace{-1cm}-[2\tilde{f}_{ml}\hspace{-.08cm}^i f^{jk}\hspace{-.08cm}_i+2\tilde{f}_{il}\hspace{-.08cm}^k f^{ij}\hspace{-.08cm}_m+\frac{1}{2}\tilde{f}_{mil} f^{jki}+2\tilde{f}_{il}\hspace{-.08cm}^k f^{ij}\hspace{-.08cm}_m+\tilde{f}_{mi}\hspace{-.08cm}^j f^{ki}\hspace{-.08cm}_l+\tilde{f}_{lm}\hspace{-.08cm}^i f^{jk}\hspace{-.08cm}_i]A^m_{\mu}A_{\nu jk}A_{\lambda {\tilde l}}+\tilde{f}_{im}\hspace{-.08cm}^l f^{ji}\hspace{-.08cm}_kA_{\mu \tilde m}^{k}A_{\nu}^{\tilde j}A_{\lambda  l}
\\
\nonumber
&&\hspace{-1cm}+[f^{lmi} f^{jk}\hspace{-.08cm}_i-f^{jli}f^{mk}\hspace{-.08cm}_i-2f^{imk}f^{jl}\hspace{-.08cm}_i+2f^{ijk} f^{lm}\hspace{-.08cm}_i{\tilde f}_{mi}\hspace{0cm}^lf^{ijk}-\frac{1}{2}f^{lmi} f^{kj}\hspace{-.08cm}_i]A_{\mu}^{\tilde k}A_{\nu j}A_{\lambda lm}+\tilde{f}_{ijk}f^{lim}A_{\mu }^{\tilde m}A_{\nu  {\tilde j}{\tilde k}}A_{\lambda m}^{\tilde l}
 \end{eqnarray}
 \begin{eqnarray}
\nonumber
&&\hspace{-1cm}-2[f^{ki}\hspace{-.08cm}_j f^{lm}\hspace{-.08cm}_i+f^{mk}\hspace{-.08cm}_i f^{il}\hspace{-.08cm}_jA_{\mu \tilde m}^{\tilde k}A_{\nu j}A_{\lambda  l}+\frac{1}{2}\tilde{f}_{ji}\hspace{-.08cm}^k f^{mli}+\frac{1}{2}\tilde{f}_{li}\hspace{-.08cm}^m f^{ik}\hspace{-.08cm}_j-\tilde{f}_{ji}\hspace{-.08cm}^l f^{mik}]A_{\mu m}A_{\nu \tilde j}^{\tilde k}A_{\lambda  l}+2f^{ki}_jf^{lm}_iA_{\mu km}A_{\nu \tilde j}A_{\lambda}^{\tilde l}
\\
\nonumber
&&\hspace{-1cm}-[f^{ml}\hspace{-.08cm}_i\tilde{f}_{jk}\hspace{-.08cm}^i-f^{mi}\hspace{-.08cm}_k\tilde{f}_{ji}\hspace{-.08cm}^l+2f^{il}\hspace{-.08cm}_k\tilde{f}_{ij}\hspace{-.08cm}^m]A_{\mu}^{\tilde m}A_{\nu {\tilde j}{\tilde k}}A_{\lambda l}+3[-\frac{1}{2}\tilde{f}_{jik}\tilde{f}_{lm}\hspace{0cm}^i+\frac{2}{3}\epsilon^{\mu\nu\lambda}\tilde{f}_{kl}\hspace{-.08cm}^i\tilde{f}_{jim}]A_{\mu \tilde k}A_{\nu}^{j}A_{\lambda {\tilde l}{\tilde m}}-\tilde{f}_{ik}\hspace{-.08cm}^j f^{il}\hspace{-.08cm}_mA_{\mu }^{\tilde k}A_{\nu}^{\tilde j}A_{\lambda  l\tilde m}
\\
\nonumber
&&\hspace{-1cm}+[\tilde{f}_{jk}\hspace{0cm}^i\tilde{f}_{im}\hspace{0cm}^l+2\tilde{f}_{ji}\hspace{-.08cm}^l{\tilde f}_{km}\hspace{-.08cm}^i+2\tilde{f}_{mik}f^{il}\hspace{-.08cm}_j+\tilde{f}_{ijk}f^{li}\hspace{-.08cm}_m]A_{\mu}^mA_{\nu {\tilde j}{\tilde k}}A_{\lambda l}
+[\tilde{f}_{ml}\hspace{-.08cm}^if^{jk}\hspace{-.08cm}_i-\tilde{f}_{li}\hspace{-.08cm}^kf^{ij}\hspace{-.08cm}_m-2f^{ij}\hspace{-.08cm}_l\tilde{f}_{im}\hspace{-.08cm}^kA-2f^{ijk}\tilde{f}_{mli}]A_{\mu k}A_{\nu j}A_{\lambda \tilde l}^m
\\
\nonumber
&&\hspace{-1cm}-[3\tilde{f}_{ik}\hspace{-.08cm}^j\tilde{f}_{ml}\hspace{-.08cm}^i+\tilde{f}_{im}\hspace{-.08cm}^j\tilde{f}_{kli}+2\tilde{f}_{ilk}f^{ji}\hspace{0cm}_m]A_{\mu {\tilde k}{\tilde m}}A_{\nu}^{\tilde j}A_{\lambda \tilde l}
-2[2\tilde{f}_{ij}\hspace{-.08cm}^k\tilde{f}_{ml}\hspace{-.08cm}^i-f^{ki}_m{\tilde f}_{ilj}A_{\mu \tilde m}-f^{ki}_j{\tilde f}_{ilm}A_{\mu \tilde m}^{\tilde k}-\frac{1}{2}\tilde{f}_{mi}\hspace{-.08cm}^l f^{ik}_l]A_{\mu \tilde m}^{\tilde k}A_{\nu \tilde j}A_{\lambda{\tilde l}}
\\
\nonumber
&&\hspace{-1cm}-[2\tilde{f}_{ij}\hspace{-.08cm}^kf^{im}_l-3f^{ki}_j\tilde{f}_{li}\hspace{-.08cm}^m+2\tilde{f}_{il}\hspace{-.08cm}^k f^{im}_j+\tilde{f}_{lk}\hspace{-.08cm}^i f^{mj}_i]A_{\mu m}A_{\nu \tilde j}^{\tilde k}A_{\lambda \tilde l}
[\tilde{f}_{im}\hspace{-.08cm}^l\tilde{f}_{jk}\hspace{-.08cm}^i+2f^{il}\hspace{-.08cm}_j\tilde{f}_{kim}+3\tilde{f}_{mj}\hspace{-.08cm}^i\tilde{f}_{ik}\hspace{-.08cm}^l+2f^{ji}\hspace{-.08cm}_l\tilde{f}_{km}\hspace{0cm}^i]A_{\mu \tilde m}A_{\nu\tilde j}^{k}A_{\lambda l}
\\
\nonumber
&&\hspace{-1cm}-[2f^{ij}\hspace{-.08cm}_k\tilde{f}_{il}\hspace{-.08cm}^m-2f^{jm}\hspace{-.08cm}_i\tilde{f}_{kl}\hspace{-.08cm}^i-2f^{im}\hspace{-.08cm}_k\tilde{f}_{il}\hspace{-.08cm}^j+f^{im}\hspace{-.08cm}_k\tilde{f}_{ijl}-f^{kj}\hspace{-.08cm}_lf^{jm}\hspace{-.08cm}_i+2f^{ij}\hspace{-.08cm}_l\tilde{f}_{ki}\hspace{-.08cm}^m-f^{mij}\hspace{-.08cm}\tilde{f}_{kil}+{\tilde f}_{jk}\hspace{0cm}^{i}\tilde{f}_{mi}\hspace{-.08cm}^l
+f^{jmi}\tilde{f}_{kil}
\\
\nonumber
&&\hspace{-1cm}+2f^{mi}\hspace{-.08cm}_l\tilde{f}_{ik}\hspace{-.08cm}^j]A_{\mu}^kA_{\nu j}A_{\lambda {\tilde l}m}
-[2f^{ki}\hspace{-0cm}_jf^{ml}\hspace{0cm}_i+\tilde{f}_{ji}\hspace{0cm}^kf^{ilm}+2\tilde{f}_{ji}\hspace{0cm}^lf^{imk}-3f^{kil}\tilde{f}_{ij}\hspace{0cm}^m]A_{\mu}^{\tilde m}A_{\nu {\tilde j}k}A_{\lambda l}+\tilde{f}_{ki}\hspace{-.08cm}^j f^{lmi}A_{\mu m}A_{\nu j}^{k}A_{\lambda l}
\\
\nonumber
&&\hspace{-1cm}-[4\tilde{f}_{ij}\hspace{-.08cm}^k\tilde{f}_{ml}\hspace{-.08cm}^i-f^{ik}_l\tilde{f}_{imj}+f^{ki}_m\tilde{f}_{lij}A_{\mu k{\tilde m}}]A_{\mu {\tilde m}}A_{\nu {\tilde j}k}A_{\lambda}^l
+2[\tilde{f}_{jk}\hspace{-.08cm}^i\tilde{f}_{li}\hspace{-.08cm}^m+\tilde{f}_{ij}\hspace{-.08cm}^m\tilde{f}_{kl}\hspace{-.08cm}^i]A_{\mu {\tilde k}m}A_{\nu\tilde j}A_{\lambda}^l
+[2\tilde{f}_{jk}\hspace{0cm}^if^{ml}\hspace{0cm}_i+f^{im}\hspace{0cm}_j\tilde{f}_{ki}\hspace{-.08cm}^l
\\
&&\hspace{-1cm}-2\tilde{f}_{jki}\tilde{f}_{lim}+\hspace{-0.1cm}f^{li}\hspace{0cm}_k\tilde{f}_{ij}\hspace{-.08cm}^m-\hspace{-0.1cm}f^{im}\hspace{0cm}_k\tilde{f}_{ij}\hspace{-.08cm}^l]A_{\mu\tilde k}A_{\nu {\tilde j}}A_{\lambda l}^{\tilde m}
-[2f^{li}\hspace{-.08cm}_m\tilde{f}_{ij}\hspace{-.08cm}^k+\hspace{-0.1cm}{\tilde f}_{ijl}f^{ki}\hspace{-.08cm}_m-\hspace{-0.1cm}f^{ik}\hspace{-.08cm}_m\tilde{f}_{ij}\hspace{-.08cm}^l-\hspace{-0.1cm}f^{ik}\hspace{-.08cm}_j\tilde{f}_{im}\hspace{-.08cm}^l\hspace{-0.1cm}-f^{li}\hspace{0cm}_j{\tilde f}_{mi}\hspace{0cm}^k]A_{\mu k}A_{\nu \tilde j}A_{\lambda l}^m{\big)},
 \end{eqnarray}
 where $F^{ABC}$ is the structure constant of the Manin triple of Lie bialgebra (${\cal D},{\cal G},{\cal G}^{*}$)\footnote{Note that 3-Lie bialgebra isn't direct sum of $\cal A$ and ${\cal A}^*$ if so, we couldn't have relation $[T^+,T^i,T^{\tilde j}]=f^{ik}\hspace{0cm}j T^{\tilde k}$ and $[T^{\tilde +},T^{\tilde i},T^j]={\tilde f}_{ik}\hspace{0cm}^jT^{k}$ and proposition in the previous section will fail for this case. Now, we investigate this case i.e. direct sum for our model then relation (\ref{WZW1}) turn into following form:
 	\begin{eqnarray}
 	\epsilon^{\mu\nu\lambda}F^{BCD}A_{\mu B}\partial_\nu A_{\lambda CD}=f^{ij}\hspace{0cm}_kA_{\mu i}\partial_{\nu}A_{\lambda j}^k+\tilde{f}_{ij}\hspace{0cm}^k A_{\mu \tilde{i}}\partial_{\nu}A_{\lambda{\tilde j}{\tilde k}}
 	\end{eqnarray}
 	and relation (\ref{WZW2}) as follows:
 	\begin{eqnarray}
\hspace{-.3cm}\epsilon^{\mu\nu\lambda}\hspace{-.09cm}F^{ABC}\hspace{-.08cm} F^{EF}\hspace{-.08cm}_A A_{\mu EF}A_{\nu B}A_{\lambda C}\hspace{-.3cm}=\epsilon^{\mu\nu\lambda}f^{jk}\hspace{-.08cm}_if^{il}\hspace{-.08cm}_m A_{\mu l}^m A_{\nu j}A_{\lambda k}+\epsilon^{\mu\nu\lambda}\tilde{f}_{ki}\hspace{-.08cm}^j\tilde{f}_{ml}\hspace{-.08cm}^iA_{\mu\tilde m}^lA_{\nu j}A_{\lambda \tilde k}
 	\end{eqnarray}
 	the result is the same one in  Ref.\cite{Mat} with one diference that we will have two Yang-Mills action one for $\cal A$ and the other for ${\cal A}^*$.}. Note that indices $A$ and $\cal A$ can be $i$ and $\tilde i$.  In the above relations we have used the notations  $A_{\mu +B}=A_{\mu {\tilde +}B}=A_{\mu B}$, $A_{\mu +{\tilde B}}=A_{\mu {\tilde +}{\tilde B}}=A_{\mu {\tilde B}}$, $F^{AB}\hspace{0cm}_C A_{\mu AB}\equiv C_{\mu C}$ and $F^{{\cal A}{\cal B}}\hspace{0cm}_{\cal C} A_{\mu {\cal A}{\cal B} }\equiv {\tilde C}_{\mu \cal C}$  that ${\cal A}=i,{\tilde i}$ and they can't  choose  only from $\cal G$ or ${\cal G}^*$, then the sum of (\ref{CS11}) and (\ref {CS22}) will have the following form:
\begin{eqnarray}
\frac{1}{2}\epsilon^{\mu\nu\lambda}\{C_{\mu B}(\partial_{\nu}A_{\lambda }^{B}-\partial_{\lambda}A_{\nu}^{B}-[A_{\nu },A_{\lambda }]_{B} )+\tilde{C}_{\mu }^{\cal B}(\partial_{\nu}A_{\lambda {\cal B}}-\partial_{\lambda}A_{\nu {\cal B}}-[A_{\nu  },A_{\lambda}]_{\cal B})\}.
\end{eqnarray}
In this way the general form of the  BLG Lagrangian on the especial 3-Lie algebra (Manin triple) (\ref{123})  is as follows \footnote{Note that in this Lagrangian the $E$ term (also "+..." terms) according to relations (44-46), can not contribute in Yang-Mills and DBI actions.}:
\begin{eqnarray}
L =  \frac{1}{2}D_{\mu}X^{A(I)}D^{\mu}X_{A}^{(I)}-2 g_{YM}^2 C_{\mu}^{B}C^{\mu}_B -2 g_{YM}{\tilde C}_{\mu {\cal B}}
  D^{\mu} X^{(8){\cal B}} + 2\,\epsilon^{\mu\nu\lambda}\,
  C_{\mu A} B_{\nu\lambda}^A +  2\,\epsilon^{\mu\nu\lambda}\,
    {\tilde C}_{\mu}^{\cal B} F_{\nu\lambda \cal B}+E
\label{L2}
\end{eqnarray}
\begin{eqnarray}
E&=&g_{YM}C^{\mu i}{\partial}_{\mu}X^{(I)}_i+g_{YM}C_{\mu i}{\partial}^{\mu}X^{i(I)}+C_{\mu i}{\tilde C}^{\mu i}-2g_{YM}f^{ik}\hspace{0cm}^jA_{\mu k}X^{j(I)}C_{\mu i}+g_{YM}{\tilde f}_{jk}\hspace{0cm}^iA_{\mu \tilde k}C_{\mu i}
\\
\nonumber
&+&g_{YM}f^{ij}\hspace{0cm}_kA_{\mu \tilde k}X^{{\tilde j}(I)}C_{\mu i}+g_{YM}f^{ikj}A_{\mu k}X^{{\tilde j}(I)}C_{\mu i}+....
\end{eqnarray}
where
\begin{eqnarray}
\label{B}
B_{\nu\lambda A}&=&\partial_{\nu}A_{\lambda A}-\partial_{\lambda}A_{\nu A}-[A_{\nu},A_{\lambda}]_A,
\\
F_{\nu\lambda }^{\cal A}&=&\partial_{\nu}A_{\lambda}^{\cal A}-\partial_{\lambda}A_{\nu}^{\cal A}-[A_{\nu},A_{\lambda}]^{\cal A}.
\end{eqnarray}
Now,  by  integration of $C_{\mu k}$ and $ \tilde{C}_{\mu}^k$  
\begin{eqnarray}
C_{\mu A}=\frac{1}{{g_{YM}}^2}\epsilon^{\nu\lambda}_{\mu}B_{\nu\lambda A}+\frac{1}{{g_{YM}}}{\partial}_{\mu}X^{(I)}_A+...
\end{eqnarray}
and 
\begin{eqnarray}
\tilde{C}_{\mu {\cal A}}=\frac{1}{{g_{YM}}^2}\epsilon^{\nu\lambda}_{\mu}F_{\nu\lambda {\cal A}}+\frac{1}{{g_{YM}}}{\partial}_{\mu}X^{(I)}_{\cal A}+...
\end{eqnarray}
then by  insertion in the Lagrangian we will obtain the following equation:
\begin{equation}
\label{FB}
L=\frac{1}{2}F_{\nu\lambda {\cal A}}F^{\nu\lambda {\cal A}}+\frac{1}{2}B_{\nu\lambda A}B^{\nu\lambda A}+\frac{1}{2}D_{\mu}X^{A(I)}D^{\mu}X_{A}^{(I)}+...,  
\end{equation}
 where $F_{\nu\lambda {\cal A}}$ is  field strength of Yang-Mills (with the gauge field $ A_{\lambda A} $) and $B_{\nu\lambda A}$ is related to B-field of a string. As we know, the dynamic of D-branes where are  expressed by DBI  and    Yang-Mills action can be obtained by expanding DBI action \cite{Tong,Pol}. Furthermore, relation between DBI action and sigma model have been investigated by Leigh in Ref. \cite{Lei} so the relation between DBI action and WZW models that are sigma model on Lie groups can be exist. In this way, our claim might be true that expanding of DBI action have a term that  correspond to a B-field of a  WZW model. We must consider that fields are 3-Lie algebraic valued in this  WZW model  that we will obtain its form  in the next subsection. Therefore,  we will have WZW model from BLG model which have been constructed on Manin triple otherwise we will have Yang-Mills model from it \cite{Mat}. In this way, we provide a method to obtain D2 from M2 and vice versa. Note that this method for obtaining D2 from M2 is different from method of  Ref.\cite{Muk2}. In the mentioned reference one can not obtain  M2 from D2 but in our method this is possible.  The BLG model is maximally supersymmetric ($N=8$) in $2+1$ dimension, with correlating DBI action being $N=(4,4)$ supersymmetric or a string $N=(4,4)$ supersymmetric. ٌWe know that if  string propagates on a group manifold, one can replace the string action with WZW action. Therefore, if we assume that our string model propagates on Lie group, then we will have an $N=(4,4)$ supersymmetric WZW model in two dimension. In the previous work we analyzed the algebraic structure of $N=(4,4)$ supersymmetric WZW model and showed that this model has Lie  bialgebra structure with one 2-cocycls \cite{AR}. Therefore, Lie algebra $\cal G$ in (\ref{example 3-bracket})  must have a Lie bialgebraic structure with 2-cocycles. In this way by starting with an $N=(4,4)$ WZW model (D2-model) with Lie algebra $\cal G$ (where it is a Lie bialgebra with one 2-cocycle) one can obtain a BLG model (M2-model) by 3-Lie algebra with commutation relation (\ref{123}) which is obtained from $\cal G$ and its dual $\cal G^*$. Note that, contrary to the ordinary WZW model, here in this model the  B-field has algebraic index and therefore we have $N=(4,4)$ like WZW model  (the form of B-field has been shown  in (\ref{B})). 

\subsection{WZW model with 3-Lie algebra valued fields}
We know that the WZW action has following form \cite{Witten}:
\begin{eqnarray}
S_{\small{WZW}}=\int d^{3}x\epsilon^{\alpha\beta\gamma}L_{\mu}\hspace{0cm}^{I}L_{\nu}\hspace{0cm}^{J}L_{\lambda}^{K}\partial_{\alpha}X^{\mu}\partial_{\beta}X^{\nu}\partial_{\gamma}X^{\lambda} Tr([T_{I},T_{J}],T_{K})
\end{eqnarray}
Now by setting the algebraic index for space-time coordinate ($X^{\mu}=X^{\mu A}T_{A}$) one can write the  WZW-like term with the following form:
\begin{eqnarray}
S_{WZW-like}=\int d^{3}x\epsilon^{\alpha\beta\gamma}L_{\mu}\hspace{0cm}^{L}L_{\nu}\hspace{0cm}^{M}L_{\lambda}^{N}\partial_{\alpha}X^{I\mu}\partial_{\beta}X^{J\nu}\partial_{\gamma}X^{K\lambda} Tr([T_{I}T_{L},T_{J}T_{M}],T_{K}T_{N}),
 \end{eqnarray}
subsequently, we anticipate that the B-field of  the WZW-like model have two algebraic indices as the following form:
\begin{eqnarray}
\nonumber
S_{WZW-like}&=&\int
d^{2}x\ \{\frac{1}{6}\epsilon^{\beta\gamma}{B_{\nu\mu}}^{Q}\partial_{\beta}X^{J\nu}\partial_{\gamma}X^{I\mu}Tr(T_{J}T_{I}T_{Q})
+\frac{1}{6}\epsilon^{\alpha\gamma}{B_{\nu\lambda}}^{Q}\partial_{\alpha}X^{J\nu}\partial_{\gamma}X^{K\lambda}Tr(T_{J}T_{K}T_{Q})
\\
&+&\frac{1}{6}\epsilon^{\alpha\gamma}{B_{\mu\lambda}}^{Q}\partial_{\alpha}X^{I\mu}\partial_{\gamma}X^{K\lambda}Tr(T_{I}T_{K}T_{Q})\}+... ,
\label{WZW  action noncordinate basis}
\end{eqnarray}
where  ${B_{\nu\mu}}^{Q}={L_{\nu}}\hspace{0cm}^{L}{L_{\lambda}}\hspace{0cm}^{N}{f_{NL}}^{P}x^J {f_{PJ}}^{Q}$, $L_{\mu}^LX^{I\mu}T_IT_L|_{boundry}=x^LT_L|_{boundry}$ such that the ${B_{\nu\mu}}^{Q}$ have the form of \ref{B}, in this way the kinetic term for WZW-like action has the  following form:
\begin{eqnarray}
\int L_{\mu}^LL_{\nu}^M \partial X^{I\mu}\partial X^{J\nu} f_{LM}^Q Tr(T_QT_IT_J).
\end{eqnarray}
Therefore, we see that the second term of (\ref{FB}) is related to the WZW-like action , i.e., a $N=(4,4)$ string (D2-model) with the Lie algebra $
\cal G$ as a Manin triple of Lie bialgebra with one 2-cocycle. This means that if we have an $N=(4,4)$ WZW model (D2-model) with Lie algebra $\cal G$ (as a Manin triple of a Lie bialgebra with one 2-cocycle) we will have a BLG model M2-model)  related to  3-Lie algebra \cite{Tong} (which is obtained from $\cal G$ and $\cal G^*$) and vice versa .
\section*{ Conclusions}
Using the concept of 3-Lie bialgebra (recently defined in arXiv:1604.04475) we have constructed BLG model on a Manin triple of a especial 3-Lie bialgebra $({\cal D},{\cal A},{\cal A}^*)$. Then, using the correspondence between 3-Lie bialgebra $({\cal A},{\cal A}^*)$ and Lie bialgebra $({\cal G},{\cal G}^*)$ we have shown that the BLG model can be reduce to an $N=(4,4)$ WZW model on a Lie algebra $\cal G$ such that the Lie algebra had a one 2-cocycle.
\\
In this way one can begin with a D2-models  with Lie algebra $\cal G$ and construst M2-model  over $\cal D$  and vice versa. One of the open problems is that one can classify such M2-models by using the classification of Lie bialgebras $\cal G$.
\section*{ Acknowledgments}
We would like to thank  M. Akbari-Moghanjoughi  for carefully reading the manuscript. This research was supported by a research fund
No. 217D4310 of Azarbaijan Shahid Madani university.
\section*{ Appendix }
 {\bf The adjoint representation of the 3-Lie algebra and ad-invariant metric on ${\cal D}$}
\\
Here we assume that $\cal D $ of Manin triple $ (\cal D,\cal A,\cal A^*) $ is a 3-Lie algebra.
\par
A Manin triple  $ (D,\cal A,\cal A^*) $which is in  one-to-one correspondence with 3-Lie bialgebras  $ (\cal A,\cal A^*) $ must  have  a nondegenerate ad-invariant inner product over 3-Lie algebra $\cal D$. In order to calculate this we choose the basis of  ${\cal D}={ \cal A}\oplus { \cal A}^*$ as; $T^A(\{T^{-},T^{+},T^i,\tilde{T}_{-},\tilde{T}_{+},\tilde{T}_i\})=\{T^-,T^+,T^i,T^{\tilde{-}},T^{\tilde{+}},T^{\tilde{i}}\}$ and use the commutation relation on $\cal A$ and $\cal A^*$:
\begin{eqnarray}
\nonumber
[T^{-},T^{a},T^{b}]=0,\hspace{1cm}
[T^{+},T^{i},T^{j}]={f}^{ij}\hspace{0cm}_{k}T^{k},\hspace{1cm}
[T^{i},T^{j},T^{k}]=f^{ijk}T^{-},
\end{eqnarray}
\begin{eqnarray}
\nonumber
[\tilde{T}_{-},\tilde{T}_{a},\tilde{T}_{b}]=0,\hspace{1cm}
[\tilde{T}_{+},\tilde{T}_{i},\tilde{T}_{j}]={\tilde{f}}_{ij}\hspace{0cm}^{k}\tilde{T}_{k},\hspace{1cm}
[\tilde{T}_{i},\tilde{T}_{j},\tilde{T}_{k}]=\tilde{f}_{ijk}T_{-},
\end{eqnarray}
Now, commutation relation for  the 3-Lie algebras $\cal D$ have the following form:
\begin{eqnarray}
[T^A,T^B,T^C]=F^{ABC}\hspace{0cm}_DT^D.
\end{eqnarray}
which we use for simplification:
\begin{eqnarray}
\left(\begin{tabular}{cccc}
        $a$ & $b$ & $c$ & $d$ \\
     $  e$ &$ f $&$ g$ &$ h$ \\
      $ i$ &$j$ &$k$ &$l $\\
     $  m $&$n$ &$ p$ &$q$ \\
      \end{tabular}\right),
\end{eqnarray}
where $a,c,i,k$ are $2 \times 2$ matrices, $b,d,j,l$ are $2 \times n$ matrices, $e,g,m,p$ are $n \times 2$ matrices and  $f,h,n,q$ are $n \times n$ matrices. Then, the adjoint representation of basis of 3-Lie algebra $\cal D$ have the following form:
\begin{eqnarray}
\nonumber
({\cal Y}^{ij})^c\hspace{0cm}_d&=&\left(\begin{tabular}{cccc}
        0 &$ b_1$ & 0 & 0 \\
      $ e_1$ & 0 & 0 & 0 \\
       0 &0 &0 &$l_1$ \\
       0 & 0&$ p_1$ &0 \\
      \end{tabular}\right),\hspace{1cm}
({\cal Y}^{\tilde{i}\tilde{j}})^c\hspace{0cm}_d=\left(\begin{tabular}{cccc}
        0 & $b_2$ & 0 & 0 \\
      $ e_2$ & 0 & 0 & 0 \\
       0 &0 &0 &$l_2$ \\
       0 & 0&$ p_2$ &0 \\
      \end{tabular}\right),
\\
({\cal Y}^{+i})^c\hspace{0cm}_d&=& \left(\begin{tabular}{cccc}
          0 &0 &0 &0 \\
       0 &$ f_1$ & 0 & 0 \\
      0 &0 &0 &0 \\
       0 &0 & 0 &$q_1$ \\
      \end{tabular}\right),\hspace{1cm}
({{\cal Y}}^{\tilde{+}\tilde{i}})^c\hspace{0cm}_d=\left(\begin{tabular}{cccc}
        0 &0 &0 &0 \\
       0 & $f_2$ & 0 & 0 \\
      0 &0 &0 &0 \\
       0 &0 & 0 &$q_2$ \\
      \end{tabular}\right),
\end{eqnarray}
and we have,  ${\cal Y}^{-i}=0 $, $ {\cal Y}^{\tilde{-}\tilde{i}}=0 $, ${\cal Y}^{-A}=0,{\cal Y}^{\tilde{-}A}=0 $, where
\begin{eqnarray}
\nonumber
b_1&=&\left(\begin{array}{ccc}
  0 & \cdots & 0 \\
  f^{ij}\hspace{0cm}_1 & \cdots & f^{ij}\hspace{0cm}_n
\end{array}\right), e_1=\left(\begin{array}{cc}
  f^{ij1} &  0 \\
  \vdots & \vdots \\
 f^{ijn} & 0
\end{array}\right),l_1=\left(\begin{array}{ccc}
  f^{ij1} & \cdots & f^{ijn} \\
 0 & \cdots & 0
\end{array}\right), p_1=\left(\begin{array}{cc}
 0 & f^{ij}\hspace{0cm}_1 \\
  \vdots & \vdots \\
0 &  f^{ij}\hspace{0cm}_n
\end{array}\right),
\\
\nonumber
b_2&=&\left(\begin{array}{ccc}
  \tilde{f}_{ij1} & \cdots & \tilde{f}_{ijn} \\
0 & \cdots & 0
\end{array}\right), e_2=\left(\begin{array}{cc}
 0&  \tilde{f}_{ij}\hspace{0cm}^1 \\
  \vdots & \vdots \\
0 & \tilde{f}_{ij}\hspace{0cm}^n
\end{array}\right),l_2=\left(\begin{array}{ccc}
 0 & \cdots &0 \\
\tilde{f}_{ij}\hspace{0cm}^1 & \cdots & \tilde{f}_{ij}\hspace{0cm}^n
\end{array}\right), p_2=\left(\begin{array}{cc}
\tilde{ f}_{ij1} & 0\\
  \vdots & \vdots \\
 \tilde{f}_{ijn} & 0
\end{array}\right),\\\
\nonumber
f_1&=&\left(\begin{array}{ccc}
  f^{i1}\hspace{0cm}_1 & \cdots &f^{i1}\hspace{0cm}_n \\
\vdots & \vdots & \vdots \\
 f^{in}\hspace{0cm}_1 & \cdots & f^{in}\hspace{0cm}_n  \\
\end{array}\right),q_1=\left(\begin{array}{ccc}
  f^{i1}\hspace{0cm}_1 & \cdots &f^{in}\hspace{0cm}_1 \\
\vdots & \vdots & \vdots \\
f^{i1}\hspace{0cm}_n & \cdots &f^{in}\hspace{0cm}_n  \\
\end{array}\right),\\
f_2&=&\left(\begin{array}{ccc}
  \tilde{f}_{i1}\hspace{0cm}^1 & \cdots &\tilde{f}_{in}\hspace{0cm}^1 \\
\vdots & \vdots & \vdots \\
 \tilde{f}_{i1}\hspace{0cm}^n & \cdots &  \tilde{f}_{in}\hspace{0cm}^n  \\
\end{array}\right),q_2=\left(\begin{array}{ccc}
  \tilde{f}_{i1}\hspace{0cm}^1 & \cdots & \tilde{f}_{i1}\hspace{0cm}^n \\
\vdots & \vdots & \vdots \\
 \tilde{f}_{in}\hspace{0cm}^1 & \cdots &\tilde{f}_{in}\hspace{0cm}^n  \\
\end{array}\right),
\end{eqnarray}
 where, $f^{ij}\hspace{0cm}_k=-({\cal X}^i)^j\hspace{0cm}_k$and $\tilde{f}_{ij}\hspace{0cm}^k=-(\tilde{\cal X}_i)_j\hspace{0cm}^k$. The non-degenerate ad-invariant inner product for Lie algebras is the result of applying the trace of bilinear product in adjoint representation \cite{Jafarizadeh}. However, if the Lie algebra is non-semisimple, current method is not useful. Obtaining  the nondegenerate ad-invariant metric for these algebras results from solving the following equation:
\begin{eqnarray}
f^{AB}\hspace{0cm}_D G^{CD}=-f^{AC}\hspace{0cm}_DG^{DB}.
\end{eqnarray}
Now, we generalize above result for 3-Lie algebra $\cal D$ and choose the basis of  ${\cal D}={ \cal A}\oplus { \cal A}^*$ as;
\\
 $T^A(\{T^{-},T^{+},T^i,\tilde{T}_{-},\tilde{T}_{+},\tilde{T}_i\})=\{T^-,T^+,T^i,T^{\tilde{-}},T^{\tilde{+}},T^{\tilde{i}}\}$, to obtain the inner product as follows:
 \begin{eqnarray}
 \nonumber
 \langle T^{A},[T^B ,T^C, T^{D}]_{{\cal D}}\rangle &=&-
 \langle[T^{A},T^B,T^C]_{{\cal D}},T^{D} \rangle
 \\
 \label{metric}
 F^{BCD}\hspace{0cm}_E G^{EA}&=& -F^{ABC}\hspace{0cm}_{E} G^{ED},
 \end{eqnarray}
  By choosing $ ({\cal Y}^{AB})^C\hspace{0cm}_D=F^{ABC}\hspace{0cm}_D $  then (\ref{metric}) means that $ {\cal Y}^{AB} G$ must be antisymmetric. We have shown their matrix representations as  follows and have concluded that the metric must have the following form:
\begin{eqnarray}
G=\left(\begin{tabular}{cccc}
        $a$ & 0 & b &0 \\
     0 &$ 0$&0 &h \\
      $ i$ &0 &$k$ &0\\
    0&$n$ &0 &0 \\
      \end{tabular}\right),
 \end{eqnarray}
where
\begin{eqnarray}
a=\left(\begin{array}{cc}
0 & g_{11}\\
 g_{12} & 0
\end{array}\right),b=\left(\begin{array}{cc}
0 & g_{21}\\
g_{22} & 0
\end{array}\right),i=\left(\begin{array}{cc}
0 & g_{31}\\
 g_{32} & 0
\end{array}\right),k=\left(\begin{array}{cc}
0 & g_{41}\\
g_{42} & 0
\end{array}\right),
\end{eqnarray}
in which  matrices $h$ and $n$ are arbitrary matrices.

\end{document}